\documentclass[epj,draft]{svjour}
\newcommand{\e}{\mbox{e}}
\newcommand{\be}{\begin{equation}}
\newcommand{\ee}{\end{equation}}
\newcommand{\bear}{\begin{eqnarray}}
\newcommand{\ear}{\end{eqnarray}}
\newcommand{\Tr}{{\rm Tr}}
\newcommand{\rvec}[1]{|#1)}
\newcommand{\lvec}[1]{(#1|}
\newcommand{\scalprod}[2]{(#1|#2)}
\newcommand{\Rvec}[1]{\Big|#1\Big)}

\newcommand{\half}{\mbox{$\frac12$}}
\renewcommand{\d}{\mbox{d}}
\newcommand{\bm}[1]{\mbox{\boldmath$#1$}}

\newcommand{\toto}{\mathop{\longrightarrow}}
\newcommand{\av}[1]{\left\langle{#1}\right\rangle}
\newcommand{\bel}[1]{ 
            \begin{equation}\label{#1}}
\newcommand{\bearl}[1]{ 
            \begin{eqnarray}\label{#1}}
\newcommand{\req}[1]{(\ref{#1})}

\begin{document}

\title {Low frequency, low temperature properties of the spin-boson problem}

\author{Heinz Horner}

\institute{ Institut f\"ur Theoretische Physik, Universit\"at Heidelberg 
Philosophenweg 19, D-69120 Heidelberg, Germany}
\date{Received: July 6, 2000 / Revised version: August 8, 2000}

\abstract
{Low temperature and low frequency properties of a spin-boson model are investigated
within a super operator and Liouville space formulation. The leading contributions
are identified with the help of projection operators projecting onto the equilibrium
state. The quantities of interest are expressed in terms of weighted bath propagators
and static linear and nonlinear susceptibilities. In particular the generalized Shiba
relation and Wilson ratio are recovered.
\PACS{
      {05.30.-d}{Quantum statistical mechanics}   \and
      {66.35.+a}{Quantum tunneling of defects}
     }}

\dedication{Dedicated to Franz Wegner on the occasion of his 60th birthday}

\maketitle


\section{Introduction}

The spin-boson model is one of the simplest models to study the interplay of quantum
mechanics and dissipation due to the interaction with an environment. It consists in a
quantum mechanical spin $\half$ weakly coupled to a macroscopic heat bath made up of
harmonic oscillators. This and similar models of dissipative quantum systems have
been widely studied in the literature \cite{CaldeiraLeggett,LeggettChakravarty,Weiss}.
 
The total system including the bath is described by a Hermitean Hamiltonian and the
corresponding Schr\"oding\-er or Heisenberg equation. The goal is, however, to obtain an
effective description of the dynamics of the subsystem having averaged over the bath
degrees of freedom. An example is the dynamics of a spin $\half$ ruled by Bloch
equations.

Especially for low frequencies and low temperatures one expects that an effective
description might exist which is of the same structure as perturbation theory but with
renormalized parameters. This is expected in analogy to the Landau-Fermi-liquid
theory where effective masses and interactions can be obtained from static
expectation values or experiments.

For the spin-boson model results of this kind are known. For instance the
ratio of the specific heat at low temperature and the static transverse susceptibility
is determined by a bath propagator weighted with the spin-bath interaction. This has
been found by Sassetti and Weiss \cite{SassettiWeiss} as generalization of the
corresponding ratio for the Kondo problem, known as Wilson ratio. Perturbation theory
yields the same result but with the transverse susceptibility replaced by its free
value. Another example is the Shiba relation for the Anderson model
\cite{Shiba} generalized again to the spin-boson model \cite{SassettiWeiss}. It
states that the long time behavior of the transverse correlation function at zero
temperature is again given by the static transverse susceptibility and the long time
property of a weighted bath propagator. Again this result agrees with perturbation
theory provided the susceptibility is replaced by its free value. 

The present paper deals with the low frequency and low temperature properties of a
spin-boson system. In particular a setup is investigated which allows to express the
low frequency and low temperature properties in terms of bath propagators and static
linear and nonlinear susceptibilities. This setup is based on a Liouville space
formulation \cite{FickSauermann,Horner}. The fist section gives a brief survey of
this formalism. The spin-boson model is introduced in section 2 and the weighted bath
propagators are discussed. Section 3 deals with the long time behavior of transverse
and longitudinal correlation and response functions. This includes the Shiba relation
and fluctuation dissipation theorems. The specific heat at low temperature and the
Wilson ratio are discussed in section 4. As mentioned above the present analysis is
based on the existence of certain static linear and nonlinear susceptibilities. The
question under which conditions they do exist or do not exist \cite{Weiss}, for
instance because of localization, is not addressed. Furthermore the susceptibilities
may depend on the high frequency cutoff of the bath oscillators resulting in non
universal prefactors.


\section{Liouville space and super operators}

A convenient starting point is a formulation in Liouville space
\cite{FickSauermann,Horner}. This is the linear space spanned by quantum mechanical
Hermitean operators and each operator $\hat A$  is considered as a vector $\rvec{A}$
in this space. Super operators $\bm{\cal O}$ are introduced as mappings of the quantum
mechanical operators onto themselves. An example is the von Neumann equation for the
statistical operator $\hat \rho(t)$
\bel{uwhs}
\frac{\d}{\d t}\hat\rho(t)=-i\Big[\hat H(t),\hat\rho(t)\Big]
\Leftrightarrow \bm{\cal L}(t)\,\Rvec{\rho(t)}
\ee
defining the Liouville super operator $\bm{\cal L}(t)$. Here and in the following
$\hbar=1$ is assumed.

There are several possibilities to define a scalar product. In the following
\be
\scalprod{A}{B}=\Tr\,\hat A \hat B
\ee
is used.

The temporal evolution is described by the super operator $\bm{\cal U} (t,t')$ which
obeys
\be
\frac{\d}{\d t}\bm{\cal U} (t,t')=\bm{\cal L} (t) \bm{\cal U} (t,t') 
\ee
with initial condition $\bm{\cal U} (t,t)=\bm{I}$ and $\bm{I}\rvec{A}=\rvec{A}$. 

The measurement of an observable $\hat A$ can also be represented as the action of a
super operator
$\bm{\cal A}$ with
\be
\bm{\cal A}\,\rvec{X}\Leftrightarrow\half \Big\{\hat A \hat X+\hat X\hat A\Big\}
\ee
where the identity vector $\rvec{1}$ obeys $\bm{\cal A} \rvec{1}=\rvec{A}$. This allows
to write the equilibrium correlation functions for time independent $\bm{\cal L}$ and
$t>t'$ as
\bel{trig}
C_{AB}(t,t')=\lvec{1}\bm{\cal A}\,\bm{\cal U}(t,t')\bm{\cal B}\rvec{\bar\rho}.
\ee
The equilibrium statistical operator obeys $\bm{\cal L}\rvec{\bar\rho}=0$.

Adding external fields to the Hamiltonian $\hat H(t)\to \hat H-h_B(t) \,\hat B$ response
functions are defined as
\bel{deur}
G_{AB}(t,t')=\frac{\delta\,\langle \hat A(t)\rangle}{\delta \,h_B(t')}
=\lvec{1}\bm{\cal A}\,\bm{\cal U}(t,t')\,\tilde{\bm{\cal B}}\,\rvec{\bar\rho}
\ee 
with
\be
\tilde{\bm{\cal B}}\,\rvec{X}\Leftrightarrow i\,\Big[\hat B,\hat X\Big].
\ee

For spin  $\half$ a complete set of quantum mechanical operators is formed by the
identity and the Pauli matrices. They can be used as basis vectors of the corresponding
Liouville space:
\bear
\rvec{1}\Leftrightarrow \left(\begin{array}{cc} \,1\,&\,0\,\\0&1\end{array}
\right)&\qquad\quad&
\rvec{x}\Leftrightarrow \left(\begin{array}{cc} \,0\,&\,\,\,1\,\\ 1&\,0\end{array}
\right) \nonumber \\
\rvec{y}\Leftrightarrow \left(\begin{array}{cc} \,0& -i\\ i&0\end{array}
\right)&\qquad\quad&
\rvec{z}\Leftrightarrow \left(\begin{array}{cc} \,1\,&\,\,0\,\\
0&-1\end{array}\right).
\ear
This results in a 4-dimensional representation
\bear
\rvec{1}\Leftrightarrow \left(\begin{array}{c}1\\ 0\\ 0\\ 0 \end{array} \right) &&\qquad
\rvec{x}\Leftrightarrow \left(\begin{array}{c}0\\ 1\\ 0\\ 0 \end{array} \right)
\nonumber\\
\rvec{y}\Leftrightarrow \left(\begin{array}{c}0\\ 0\\ 1\\ 0 \end{array} \right) &&\qquad
\rvec{z}\Leftrightarrow \left(\begin{array}{c}0\\ 0\\ 0\\ 1 \end{array} \right).
\ear
Super operators are then $4\times4$ matrices, for instance
\bear
{\bm\sigma}_x\Leftrightarrow 
\left(\begin{array}{cccc}\,0\,&\,1\,&\,0\,&\,0\, \\1&0&0&0\\0&0&0&0\\0&0&0&0
\end{array}\right)&\quad&
\tilde{\bm\sigma}_x\Leftrightarrow 2
\left(\begin{array}{cccc}\,0\,&\,0\,&\,0\,&\,0\, \\0&0&0&0\\0&0&0&1\\0&0&-1&0
\end{array}\right) \nonumber\\
{\bm\sigma}_z\Leftrightarrow 
\left(\begin{array}{cccc}\,0\,&\,0\,&\,0\,&\,1\, \\0&0&0&0\\0&0&0&0\\1&0&0&0
\end{array}\right)&\quad&
\tilde{\bm\sigma}_z\Leftrightarrow 2
\left(\begin{array}{cccc}\,0\,&\,0\,&\,0\,&\,0\, \\0&0&1&0\\0&-1&0&0\\0&0&0&0
\end{array}\right). 
\ear

For particles, especially the bath oscillators, the Wig\-ner representation is used. The
corresponding Liouville space is spanned by functions 
$\rvec{X}\Leftrightarrow X(p,x)$ of the coordinates $x$ and momenta $p$. The scalar
product is
\be
\scalprod{Y}{X}=\int\d p \d x\,Y(p,x)\,X(p,x).
\ee
The super operators for the measurement of position and momentum and the corresponding
response operators are
\bear
\bm{x}\Leftrightarrow x&\qquad\quad&
\tilde{\bm{x}}\Leftrightarrow-\frac{\partial}{\partial p}\nonumber\\
\bm{p}\Leftrightarrow p&\qquad\quad&
\tilde{\bm{p}}\Leftrightarrow\;\frac{\partial}{\partial x} \;.
\ear
A general super operator $\bm{\cal O}$ in Wigner representation is a function of four
variables ${\cal{O}}(x,p,x',p')$.

The Liouville space of the combined system and bath oscillators is the direct product 
of the individual spaces, for instance the four dimensional space of a spin $\half$
and a Wigner representation for each bath oscillator.


\section{Spin-boson problem}

The Hamiltonian of a tunneling center or a spin $\half$ interacting with a thermal bath
of
$N$ harmonic oscillators is
\bearl{sfre}
\hat H&=&-\half \Delta\hat\sigma_x-\half \varepsilon\hat\sigma_z
-\frac{1}{\sqrt{N}}\sum_k\Lambda_k\hat x_k\hat \sigma_z \nonumber \\
&&+\half\sum_k\Big\{\hat p_k^2+\omega_k^2\,\hat x_k^2\Big\}
\ear
where $\Delta$ is the bare tunneling splitting and $\varepsilon$ a bias or anisotropy. 
Eventually the limit $N\to\infty$ is considered and the $1/\sqrt{N}$ scaling of the
spin-bath coupling constants is explicitly written down. In the following 
the isotropic case $\varepsilon=0$ is considered primarily, but generalizations to the
general case are possible.

The resulting Liouville operator \req{uwhs} is
\bear
\bm{\cal L}&=&\half \Delta\tilde{\bm\sigma}_x+\half \varepsilon\tilde{\bm\sigma}_z
+\frac{1}{\sqrt{N}}\sum_k\Lambda_k\Big\{\bm\sigma_z\tilde{\bm x}_k
+\bm x_k\tilde{\bm\sigma}_z\Big\} \nonumber\\ 
&&-\sum_k\Big\{\bm p_k\,\tilde{\bm p}_k+\omega_k^2\bm{x}_k\tilde{\bm x}_k\Big\}.
\ear

The bath is characterized by its density of states
\bel{ieby}
J(\omega)=\frac{1}{N}\sum_k\frac{\Lambda_k^2}{\omega_k}\delta(\omega-\omega_k)=
\alpha \Theta^{1-s}\omega^se^{-\omega/\Theta}
\ee
where the special case of a power law dependence is of interest. The case $s=1$ is
usually referred to as Ohmic case. A tunneling center in a crystal or glass requires a
super Ohmic spectrum with $s=3$. There is a high frequency cutoff $\Theta$
representing the Debye frequency and the same frequency is used to define the
dimensionless coupling constant
$\alpha$  for $s\ne 1$.

Because of the $1/\sqrt{N}$ dependence of the couplings the correlation and response functions
of the bath oscillators are unperturbed. For an oscillator with frequency $\omega_k$ the
$x_k$-$x_k$-response function for $t>0$ is
\be
F_k(t)=\frac{1}{\omega_k}\sin \omega_k t
\ee
and the corresponding correlation function at temperature $T=1/\beta$ is
\be
D_k(t)=\frac{n(\omega_k)}{\omega_k}\cos \omega_k t
\ee 
with
\be
n(\omega)=\half\coth\half\beta\omega.
\ee
The action of the bath on the spin involves the weighted bath propagators
\be
F(t)=\frac{1}{N}\sum_k \Lambda_k^2 F_k(t)
\ee 
and
\be
D(t)=\frac{1}{N}\sum_k \Lambda_k^2 D_k(t).
\ee 
Introducing the Fourier transforms
\bel{kgvr}
\hat F(\omega)=\int_0^\infty \d t\,\e^{i\omega t} F(t)
\ee
and
\be
\hat D(\omega)=\int_{-\infty}^\infty \d t\,\e^{i\omega t} D(t)
\ee
one finds for the imaginary part of $\hat F(\omega)$
\be
\Im \hat F(\omega)=\frac{\pi}{2} \mbox{sign}(\omega) \, J(|\omega|)
\ee
and for $\hat D(\omega)$
\be
\hat D(\omega)=\pi n(|\omega|) J(|\omega|)=2 n(\omega)\Im \hat F(\omega).
\ee

This last expression is a special case of the quantum mechanical fluctuation dissipation theorem
(FDT) \cite{FickSauermann,Horner}
\bel{ourg}
\hat C_{AB}(\omega)=2 n(\omega) \Im \hat G_{AB}(\omega)
\ee
which holds for any pair of operators $A$ and $B$ having the same parity with
respect to time reversal.

For the power law density of states \req{ieby} one finds
\bearl{jeyc}
F(t)&=&\alpha \,\Theta^2\Gamma(s+1)\sin\Big((s+1) \arctan(\Theta t)\Big) \nonumber\\
&&\times\Big(1+(\Theta t)^2\Big)^{-(s+1)/2}
\ear
and for $T=0$
\bearl{duhw}
D_0(t)&=&\half\alpha\Theta^2\Gamma(s+1)\cos\Big((s+1)\arctan(\Theta t)\Big) 
\nonumber \\
&&\times\Big(1+(\Theta t)^2\Big)^{-(s+1)/2}.
\ear
The leading correction for finite temperature is $\delta D(t)=D(t)-D_0(t)$ with 
\bearl{xirn}
\delta D(t) &=&\int_0^\infty \d \bar\omega 
\frac{J(\bar\omega)}{\e^{\beta \bar\omega}-1}\,\cos \bar\omega t\nonumber\\
&=&\alpha\,T^2\,(\Theta/T)^{1-s}\,d(T\,t)
\ear
and
\bear
d(\tau)&=&\int_0^\infty \d x \, \frac{x^s}{\e^x-1}\,\cos x\tau \nonumber\\
&=&\sum_{\ell=1}^{\infty} \Gamma(s+1)\,\cos\Big(
(s+1)\arctan(\tau/\ell)\Big) \nonumber\\
&&\times\Big(\ell^2+\tau^2\Big)^{-(s+1)/2} \nonumber \\
&\displaystyle \toto_{\tau\to 0}& \Gamma(s+1)\zeta(s+1)
-\half\Gamma(s+3)\zeta(s+3) \tau^2 \nonumber \\
&\displaystyle \toto_{\tau\to \infty}& \Gamma(s)\cos(s\pi/2)\,\tau^{-s}
+\half\Gamma(s+1)\nonumber\\
&&\times\sin(s\pi/2)\,\tau^{-s-1}.
\ear

A formal perturbation theory for expressions like \req{trig} or \req{deur} can be set up by
writing 
\be
\bm{\cal U}(t,t')=\bm{T}\Big[ \exp {\int_{t'}^{t}\d s\,\bm{\cal L}(s)}\Big]
\ee
where $\bm T$ denotes appropriate time ordering. For the equilibrium statistical
operator
$\rvec{\bar\rho}=\bm{\cal U}(t',-\infty)\rvec{i}$ is used. The initial state is not relevant
assuming that the system has equilibrated at time $t'$. As a next step the exponential
is expanded in a sum of products of free evolution operators in spin space  and
bath operators. Finally all bath operators  have to be contracted in pairs to bath
response functions
$F(t)$ or correlation functions $D(t)$.

\section{Long time behavior}

The following analysis of the long time behavior is based on two arguments:

1) The system tends towards equilibrium

\be
\bm{\cal U}(t)\toto_{t\to\infty} \rvec{\bar\rho}\lvec{1}
\ee

2) The long time properties are ruled by the long time properties of the averaged bath
response function $F(t)$, \req{jeyc}, and the bath correlation function $D(t)$, \req{duhw} and
\req{xirn}.

In order to isolate the leading contributions at long time a set of projection operators is
introduced
\bearl{etdi}
\bm{\cal P}_0&=&\rvec{\bar\rho}\lvec{1} \\
\bm{\cal P}_1&=&\sum_k \Big\{\tilde{\bm{x}}_k\rvec{\bar\rho}\lvec{1}\bm{p}_k
-\tilde{\bm{p}}_k\rvec{\bar\rho}\lvec{1}\bm{x}_k\Big\} \nonumber \\
\bm{\cal P}_\ell&=&\frac{1}{\ell !}\,\mbox{\bf :}
\Big[\sum_k\Big\{\tilde{\bm{x}}_k {\bm{p}}_k-\tilde{\bm{p}}_k {\bm{x}}_k\Big\}
\Big]^{\ell}\rvec{\bar\rho}\lvec{1}\, \mbox{\bf :} \nonumber
\ear
where the normal product $\mbox{\bf :}\cdots \mbox{\bf :}$ means that all bath
response operators
$\tilde{\bm{x}}_k$ and $\tilde{\bm{p}}_k$ have to be arranged to the left of
$\rvec{\bar\rho}\lvec{1}$ and the operators ${\bm{x}}_k$ and ${\bm{p}}_k$ to the right.
Furthermore for $\ell>1$ each value of $k$ may appear only once. 

Assume a projector $\bm{\cal P}_\ell$ is inserted into the time evolution 
$\bm{\cal U}(t,t')$ at some time \mbox{$t>t_0>t'$}
\be
\bm{\cal U}(t,t')\to \bm{\cal U}(t,t_0) \bm{\cal P}_\ell \, \bm{\cal U}(t_0,t').
\ee
If the contraction of pairs of bath operators to bath propagators is performed, as 
outlined at the end of the previous section, exactly $\ell$ such propagators reach
from some time $>t_0$ to some time $<t_0$. The proposal is that the leading
contribution for long time is given by the lowest order projector which can be
inserted.
 
Let me start with the investigation of the response function 
\bear
G_{zz}(t)&=&\lvec{1}\bm{\sigma}_z \bm{\cal U}(t,0) \tilde{\bm{\sigma}}_z \rvec{\bar\rho}
\nonumber \\
&=&\lvec{1}\bm{\sigma}_z \bm{\cal U}(t,0) \tilde{\bm{\sigma}}_z\bm{\cal U}(0,-\infty) 
\rvec{i}.
\ear
Trying to insert $\bm{\cal P}_0$ at some time $t>t_0>0$  results in vanishing
contributions because 
$\lvec{1}\bm{\cal U}(t_0,0)=\lvec{1}$ and $\lvec{1}\tilde{\bm{\sigma}}_z=0$. 
Next $\bm{\cal P}_1$ is inserted. This yields
\bearl{tvgd}
\lvec{1}&&\!\!\bm{\sigma}_z \bm{\cal U}(t,t_0)\bm{\cal P}_1\bm{\cal U}(t_0,0)
\tilde{\bm{\sigma}}_z\bm{\cal U}(0,-\infty)\rvec{i} \nonumber \\
&&=\int_{t_0}^t \d s \int_0^{t_0}\d s' \lvec{1}\bm{\sigma}_z \bm{\cal U}(t,t_0)
\tilde{\bm{\sigma}}_z\rvec{\bar\rho} F(s-s') \nonumber \\
&&\;\;\;\;\times \lvec{1}\bm{\sigma}_z \bm{\cal U}(t_0,0)
\tilde{\bm{\sigma}}_z\rvec{\bar\rho} \nonumber \\
&&=\int_{t_0}^t \d s \int_0^{t_0}\d s' G_{zz}(t-s)  F(s-s') G_{zz}(s') .
\ear

It is proposed that \req{tvgd} yields the leading contribution to $G_{zz}(t)$ for 
$t\to\infty$ and choosing for instance $t_0=t/2$ the $s$- and $s'$-integration can be
performed replacing $F(s-s')$ by $F(t)$. This results in 
\bearl{yrfd}
G_{zz}(t)\toto_{t\to\infty}&& \bar\chi_{zz}^2 F(t)  \\
\toto_{t\to\infty}&&\alpha \Theta^{1-s} \Gamma(s+1) \cos( \half s \pi) 
\bar\chi_{zz}^2\, t^{-1-s}\nonumber
\ear
where 
\bel{othd}
\bar \chi_{zz}=2\frac{\partial \av{\sigma_z}}{\partial \varepsilon}
=\int_0^{\infty} \d t \, G_{zz}(t)
\ee 
is static susceptibility which is assumed to be finite.
This means that the asymptotic behavior of the time dependent transverse response 
function is determined by the static susceptibility and the averaged bath propagator.
The bare parameters $\Delta$, $\varepsilon$ and temperature enter only implicitly via
$\bar\chi_{zz}$. For the Ohmic case $s=1$ the first line of \req{yrfd} is still valid
but in the second line $F(t)\to 2\alpha \Theta^{-1}t^{-3}$ has to be inserted. 

There are several points which have to be checked: In the integrals of \req{tvgd}  we
have replaced $F(s-s')$ by $F(t)$ assuming that the main contribution comes from
$t-s\ll t$ and $s'\ll t$. This is actually the case for a bath spectrum with $s>0$. If
the asymptotic contribution \req{yrfd} is inserted for $G_{zz}(t-s)$ or $G_{zz}(s')$
in \req{tvgd} corrections $\sim t^{-1-2s}$ are obtained.

Instead of $\bm{\cal P}_1$ some other projector $\bm{\cal P}_\ell$ could have been 
inserted. For a symmetric system with $\varepsilon=0$ the next nonvanishing
contribution is obtained from \mbox{$\ell=3$}. A corresponding evaluation yields a
contribution $\half\bar\chi_{zzzz}^2 F(t) D^2(t)$ where
\mbox{$\bar\chi_{zzzz}=8\partial^3\av{\sigma_z}/\partial\varepsilon^3$} is a static
nonlinear susceptibility. This contribution vanishes \mbox{$\sim t^{-3(1+s)}$} for
$t\ll \beta$ or $\sim T^2 t^{-1-3s}$ for $t\gg \beta$. For $\varepsilon\ne 0$
corrections $\sim t^{-2(1+s)}$ or $\sim T t^{-1-2s}$ result from insertion of
$\bm{\cal P}_2$. In any case \req{yrfd} yields the leading contribution. 

Further contributions arise from insertion of \ 
\mbox{$\bm{\cal Q}=\bm{I}-\sum_\ell \bm{\cal P}_\ell$}. They are determined by the decay
towards equilibrium. Within perturbation theory or NIBA
\cite{CaldeiraLeggett,LeggettChakravarty,Weiss} this decay is exponential and as a
result the leading  contribution \req{yrfd} is indeed due to the first nonvanishing
insertion of the projector $\bm{\cal P}_\ell$ with lowest $\ell=1$. The above
analysis yields the restriction $s>0$. In addition the existence of the static
susceptibility \req{othd} has been assumed which may impose further restrictions
\cite{Mielke-sub} on
$s$ or $\alpha$.

The Fourier transform
\be
\hat G_{zz}(\omega)=\int_0^\infty \d t \, \e^{i\omega t}\,G_{zz}(t)
\ee
can be written as
\bearl{fgek}
\hat G_{zz}(\omega)&=&\frac{1}{\bar\chi_{zz}^{-1}-\hat \Sigma_{zz}(\omega)}\\
&\mbox{$\displaystyle \toto_{\omega\to0}$}&\bar\chi_{zz}
+\bar\chi_{zz}\Sigma_{zz}(\omega)\bar\chi_{zz}
\ear
with a self energy $\Sigma_{zz}(\omega)$ vanishing for $\omega\to 0$. With \req{kgvr}
and \req{yrfd}
\be
\hat\Sigma_{zz}(\omega) \toto_{\omega\to 0} \hat F(\omega)-\hat F(0)\sim \omega^s .
\ee

The analysis of the correlation function \ $C_{zz}(t)$ follows similar lines. The 
Fourier transform
\be
\hat C_{zz}(\omega)=\int_{-\infty}^\infty \d t \, \e^{i\omega t}\,C_{zz}(t)
\ee
may be written as
\be
\hat C_{zz}(\omega)=\hat G_{zz}(\omega) \hat \Gamma_{zz}(\omega) \hat C_{zz}(-\omega) 
\ee
defining the self energy $\Gamma_{zz}(t)$.
Again the leading contribution at low frequencies or long time is due to an insertion 
of $\bm{\cal P}_1$. Depending on whether a bath response function $F$ or a bath
correlation function  $D$ crosses the insertion point at $t_0$ several contributions
are found. In analogy to \req{tvgd}
\bearl{mudw}
C_{zz}(t)&=&\int_{t_0}^t \d s \int_{-\infty}^{t_0}\!\!\! \d s' G_{zz}(t-s) F(s-s')
C_{zz}(s') \\ 
&&+\int_{t_0}^t \d s \int_{-\infty}^{0}\!\!\!\! \d s' G_{zz}(t-s) D(s-s')
G_{zz}(-s') . \nonumber
\ear
The second term is identified as due to the long time or low frequency part of
$\Gamma_{zz}(t)$ and therefore
\be
\hat\Gamma_{zz}(\omega)\toto_{\omega\to0} \hat D(\omega) .
\ee

The bath propagators fulfill an FDT \req{ourg} and therefore the same holds for the self energy
$\Sigma_{zz}(\omega)$ and $\Gamma_{zz}(\omega)$. As a consequence the correlation- and
response functions $C_{zz}$ and $G_{zz}$ also fulfill the FDT. This is not surprising.
It shows the consistency of the above analysis.

The leading time dependence for $t\to\infty$ is in analogy to \req{yrfd}
\bel{fysb}
C_{zz}(t)\toto_{t\to\infty} \hat\chi_{zz}^2 D(t)
\ee
and with \req{duhw} $C_{zz}(t)\sim t^{-s-1}$ for $t\to\infty$ and $t\ll \beta$. In the
opposite limit  $t\to\infty$ and $t\gg \beta$ an asymptotic decay $\sim T\,t^{-s}$ is
found from \req{xirn}. 

The above result is a generalization of the Shiba relation originally proposed for
the Kondo problem \cite{Shiba} and generalized to the spin boson problem by Sassetti
and Weiss \cite{SassettiWeiss} and others \cite{Mielke,Spohn} for $T=0$. Note that their 
definition of $\hat\chi_{zz}$ differs by a factor 4 from the present one.

The analysis of the long time properties of the longitudinal response function 
\be
G_{xx}(t)=\lvec{1}\bm\sigma_x \bm{\cal U}(t)\tilde{\bm\sigma}_x\rvec{\bar\rho}
\ee
follows similar lines. The lowest order projector which can be inserted is 
$\bm{\cal P}_2$ resulting in
\bearl{dtrk}
G_{xx}(t)&\approx&\int_{t_0}^t \d s\int_{t_0}^t \d r 
\int_{0}^{t_0}\d s' \int_{-\infty}^0\d r'\,
\av{\sigma_x(t)\tilde \sigma_z(s)\tilde\sigma_z(r)} \nonumber\\
&&\times \Big[ \,F(s-s') D(r-r')\av{\sigma_z(s')\tilde\sigma_x(0)\tilde\sigma_z(r')}
\nonumber \\
&&+F(s-s') F(r-r')\av{\sigma_z(s')\tilde\sigma_x(0)\sigma_z(r')}\Big]
\ear
where $\av{\cdots}$ are correlation-response functions of higher order. Investigating
the contribution of the first term, $F(s-s')D(r-r')$ is replaced by $F(t)D(t)$
resulting in
\be
G_{xx}(t)\toto_{t\to\infty}\bar\chi_{xzz}^2 F(t)D(t)
\ee
where 
\be
\bar\chi_{xzz}=4\frac{\partial^2\av{\sigma_x}}{\partial \varepsilon^2}
=4\frac{\partial^2\av{\sigma_z}}{\partial \Delta \partial \varepsilon}
\ee
is a static nonlinear susceptibility. The second term in \req{dtrk} can be shown to
be subdominant for $t\to\infty$. As a result again the leading contribution at long
time has been expressed in terms of nonlinear susceptibilities and weighted bath
propagators. 

The longitudinal correlation function is obtained by applying fluctuation
dissipation theorems. The resulting asymptotic expression is 
\bel{hyeg}
C_{xx}(t)-\av{\sigma_x}^2\toto_{t\to\infty}\half\, \bar\chi_{xzz}^2 \,D(t)^2
\ee
decaying $\sim t^{-2-2s}$ for $t\ll\beta$ and $\sim T^2 t^{-2s}$ for $t\gg\beta$. The 
same asymptotic behavior for $T=0$ has been obtained by Guinea \cite{Guinea} for s=1.
The  result obtained by  Lang et al. \cite{Lang} can not be compared since it refers to
the longitudinal correlation function involving polaron dressed operators whereas
\req{hyeg} involves the bare operator $\sigma_z$.


\section{Specific heat}

In this section it is shown that thermal properties at low temperature can also be
expressed in terms of static susceptibilities and bath propagators. The free energy of
the system at temperature $T=1/\beta$ is
\be
F_T=-\frac{1}{\beta}\Big\{\ln\Tr\;\e^{-\beta H}-\ln\Tr\;\e^{-\beta H_B}\Big\}
\ee
where $H$ is given by \req{sfre} and $H_B$ is the Hamiltonian of the bath without
coupling to the spin.

Let me investigate the derivative
\bear
\frac{\partial F_T}{\partial\Lambda_k}
&=&-\frac{1}{\sqrt{N}}\av{x_k\,\sigma_z}\nonumber\\
&=&-\frac{1}{N}\int_0^\infty\!\!\!\!\d t\,\Lambda_k
\av{\sigma_z(t)\Big\{D_k(t)\tilde\sigma_z(0)+F_k(t)\sigma_z(0)\Big\}}\nonumber\\
&=&-\frac{1}{N}\int_0^\infty\!\!\!\!\d t
\,\Lambda_k\Big\{D_k(t)G_{zz}(t)+F_k(t)C_{zz}(t)\Big\} . 
\ear

Because of the $1/\sqrt{N}$ dependence of the spin-bath couplings the functions 
$G_{zz}(t)$ and $C_{zz}(t)$ can be treated as independent on $\Lambda_k$ for
$N\to\infty$. This means that $F$ is a linear functional of $\Lambda_k^2 D_k(t)$ and 
$\Lambda_k^2 F_k(t)$ respectively.

The free energy $F_T$ depends on temperature only via $D_k(t)$. In order to
evaluate \mbox{$\delta F_T=F_T-F_0$} the excess
\bear
\delta D_k(t)&=&D_k(t;T)-D_k(t;0) \nonumber\\
&=&\frac{1}{\omega_k}\,\frac{1}{\e^{\beta\omega_k}-1}\,\cos \omega_k t
\ear
is introduced. At low temperature the leading contribution is linear in 
$\delta D_k(t)$ and
\bearl{ugte}
\delta F_T&\mbox{$\displaystyle\toto_{T\to0}$}&-\frac1{2N}\sum_k\Lambda_k^2
\int_0^\infty\d t\, \delta D_k(t) G_{zz}(t) \\
&=&-\half \int_0^\infty\d t\, \delta D(t) G_{zz}(t) \nonumber
\ear
with $\delta D(t)$ given by \req{xirn}. For $T\to 0$ the integral in \req{ugte} can be
evaluated by using $\delta D(0)$ instead of $\delta D(t)$ because this function
actually varies on a time scale $\beta=1/T\to\infty$. This results in
\bear
\delta F_T\, &\displaystyle\toto_{T\to 0}&\, - \half \bar \chi_{zz} \delta D(0) \\
&=&-\half\bar\chi_{zz}\,\alpha\,T^2\,(\Theta/T)^{1-s}\,\Gamma(s+1)\,\zeta(s+1) .
\nonumber
\ear
The specific heat is obtained from
\be
C=-T\,\frac{\partial^2 \delta F_T}{\partial T^2}
\ee
resulting in 
\be
C=\half\bar\chi_{zz}\,\alpha\,T^s\,\Theta^{1-s}\,s\,(s+1)\,\Gamma(s+1)\,\zeta(s+1) .
\ee
which is a generalization of the Wilson ratio \cite{SassettiWeiss}. Note again  the
difference in the definition of the transverse static susceptibility.
\\

\begin{acknowledgement}
It is a pleasure to thank Reimer K\"uhn and Andreas Mielke for stimulating
discussions. Support within the EPS program SPHINX is also acknowledged.
\end{acknowledgement}



\end{document}